# ENHANCED LOCATION BASED ROUTING PROTOCOL FOR 6LOWPAN

M. REHENA SULTHANA[1], P.T.V BHUVANESWARI[2], N. RAMA[3]

[1]Department of Computer Science, Mohamed Sathak College, Chennai.
rehena_aleem@yahoo.com

[2]Department of Electronic Science, MIT Anna University, Chennai
ptvmit@annauniv.edu

[3]Department of Computer Science, Presidency College, Chennai
nramabalu@gmail.com

## ABSTRACT

*6LoWPAN (IPv6 over IEEE 802.15.4) standardized by IEEE 802.15.4 provides IP communication capability for nodes in WSN. An adaptation layer is introduced above the MAC layer to achieve header compression, fragmentation and reassembly of IP packets. The location-based information is used to simplify the routing policy. This paper proposes an efficient location-based routing protocol, considering link quality and distance between nodes as the routing metric. The proposed Enhanced Location-based routing protocol (ELBRP) was simulated in NS2 version 2.32 and performance were analysed in terms of packet delivery ratio, throughput and average end-to-end delay. From the results obtained, it is found that the proposed ELBRP outperforms existing LOAD protocol.*

## 1. INTRODUCTION

A Wireless Sensor Network (WSN) consists of small nodes with sensing, computation and wireless communication capabilities deployed either inside the phenomenon or very close to it. To morph WSN from Personal Area Network (PAN) into low power PAN (LoW PAN), IEEE 802.15.4 standard was introduced which specifies the PHY and MAC layers for low rate WPAN. Some of the existing sensor network protocols were non-IP based such as zigbee. Extending IP to LoWPANs was thought to be impractical because they are highly constrained network in terms of memory, energy etc.,

The Zigbee Alliance and the IEEE 802.15.4 task group joined together to specify a standard protocol stack for low rate wireless sensor networks. This became a standard solution for low cost, low power monitoring control devices in industrial automation [15].

IPv6, the next generation internet protocol was developed as a successor to IPv4 to increase the scalability of the internet. The IPv6 protocol was developed to solve the IPv4 address exhaustion problem, so it expands the IP address space from 32 to 128 bit. Also IPv6 increases the Minimum Transmission Unit (MTU) requirement from 576 to 1,280 bytes considering the growth in link bandwidth [1].

Internet Engineering Task Force (IETF) standardized the transmission of IPv6 over LoWPANs through a working group known as 6LoWPAN [2]. To achieve this objective, an adaptation layer with various mechanisms like header compression, packet fragmentation and reassembly has been introduced above the logical link layer of IEEE 802.15.4 protocol stack. Similar to zigbee based WSN, 6LoWPAN based WSN are resource constrained networks. Figure 1 illustrates the protocol stack for 6LoWPAN architecture[3].

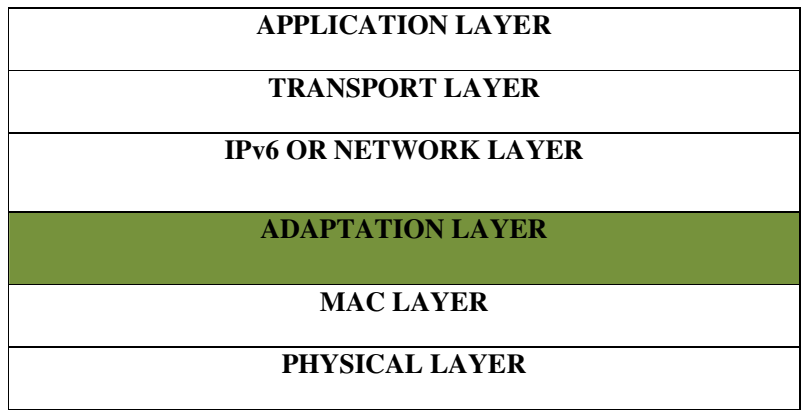

Figure 1. Protocol Stack of 6LoWPAN Architecture

Recent WSN based routing protocols are non-IP based, so to provide IP communication capability, 6LoWPAN routing protocols were developed. IP based devices can more easily be connected to other IP networks, without the need for translation gateways. Due to energy constraints, routing in WSN was performed in a multi-hop manner. Selection of best forwarding path would be the major criteria in designing a routing protocol based on 6LoWPAN.Routing in 6LoWPAN was divided in to two types as mesh-under and route-over. In mesh-under the forwarding decision is taken at the adaptation layer and in route-over the forwarding decision is taken at the network layer [4].

Based on the network structure, the routing protocols of 6LoWPAN are classified into data-centric, Hierarchical and Location-based as shown in the Figure 2. In data-centric protocols, queries on particular data are sent from base station to the network. The nodes holding the answers for the query alone send reply. It helps in maintaining many redundant transmissions. In hierarchical protocols, nodes are grouped in to clusters.

A node is elected as a cluster head (CH). CH performs aggregation of data transmitted by its cluster member using standard information fusion technique. In Location-based protocols, position information of nodes is utilised to relay the data to the desired destination. Power optimisation can be achieved in Location-based routing protocols and control overheads can be minimized.

There have been only few routing protocols developed for 6LoWPAN .They are LOAD,MLOAD,DYMO-Low, Hi-Low, S-AODV,SPN, Improved Hi-Low [5], TA-Hilow and SPEED.

The limitations of LOAD were its repeated broadcast of RREQ for route discovery process which increases the energy consumption [6]. So MLOAD [7] was proposed to find the multiple routes during the route discovery process. In MLOAD if a path fails alternate paths were used for reducing the overhead of route discovery process. The Hello messages used in DYMO-Low gave a more reliable data forwarding but resulted with higher delay in the packet routing.

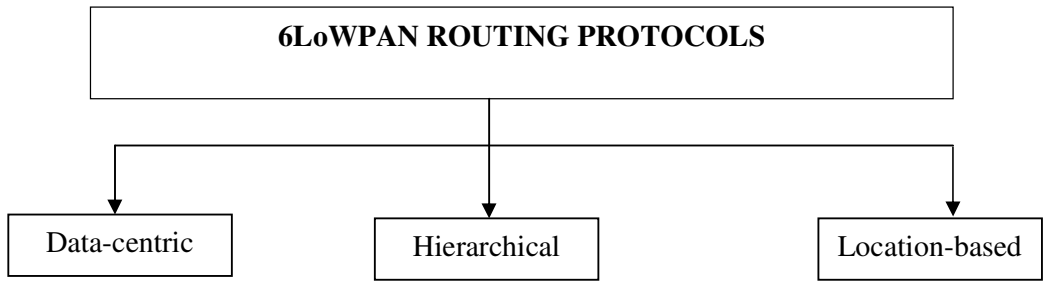

Figure 2. Taxonomy of Routing protocols

The convergence to network topology provided by Hi-Low [8] was slow even it provides a larger scalability. This will induce for more delay for the route discovery process in Hi-Low. Path recovery mechanism in conventional Hi-Low was solved by applying the SPN algorithm in Hi-Low .The SPN algorithm provides a sustainable connection along the 6LoWPAN routing path. But still performance would degrade when a node tries to communicate with other nodes over several hops. So an efficient routing scheme was required for communication between source and the sink nodes.

6LoWPAN adopts IPv6 stateless address auto configuration [12]. Any 6LoWPAN device in a network supports both 64-bit long or 16-bit short address. In IEEE 802.15.4, MAC header, Mesh addressing header and IPv6 compressed header all were related to IP address. Since IP address was important information to be carried in the 6LoWPAN network [14], addition of location-based IP address would reduce the overhead of the transmission. This provides advantages for optimisation in routing and network management. This paper proposes the idea of Location-based routing scheme in 6LoWPAN. Compared to LOAD in the proposed protocol more than one source nodes are involved in the network. Here the limitation of LOAD was overcome by adding the location based information during the broadcast of RREQ and unicast of RREP. Considering the Link quality (LQI) of the nodes reduces the routing overhead and chooses the best path based on the above parameters to provide good support for the design of routing protocols and management mechanisms.

The rest of the paper is organised as follows. Section 2 provides the back ground work in 6LoWPAN routing. In section 3, the proposed Enhanced Location-based routing protocol (ELBRP) is described in detail. In section 4, the results and discussions of proposed ELBRP is analysed and finally conclusions and future work are presented.

## 2. BACKGROUND WORK

Routing can be referred as a technique by which information are transferred from one place to another. 6LoWPAN architecture enables the transmission of IPv6 over WSNs based on the IEEE 802.15.4 standard. Various requirements of this technology are support for sleep/listen mode, low overhead on data packets, low routing overhead and minimal computation and energy requirements. The IPv6 packet size is larger than that of IEEE 802.15.4 frame. In order to accommodate IPv6 packet inside the IEEE 802.15.4 frame, an adaptation layer is introduced between the MAC and the network layer. It does the header compression, fragmentation and forwarding of packets.6LoWPAN supports routing in both layer2 and layer 3. If it happens in layer2 it is called mesh-under while in layer 3 it is called route-over.

Generally the routing protocols can be classified in to 3 categories namely proactive, reactive and hybrid based on the route discovery process. In proactive routing protocols, all the routes are computed in advance, so they are best suited for static network. In reactive routing protocols routes are computed on demand, so they are best suited for dynamic network environment .Hybrid protocols use the combination of these two protocols. Another class of routing protocol called co-operative in which sensor nodes send data to the central node which aggregates and process the data. There have been only few routing protocols developed for 6LoWPAN.And still research is going on in developing new routing protocols based on the current requirement scenario. Some of the routing protocols developed for 6LoWPAN include LOAD, DYMO-Low, Hi-Low, Improved Hi-Low, MLOAD, S-AODV, SPN, TA-HiLow (Tree Avoiding technique for hierarchical routing) and SPEED.  The classification of existing 6LoWPAN routing protocols are illustrated in Figure 3.

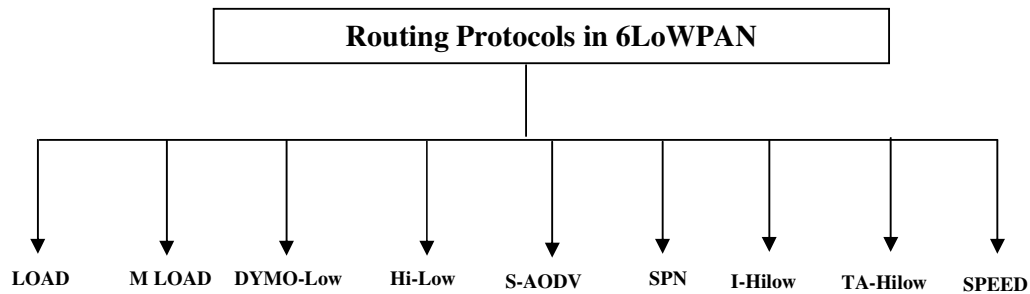

Figure 3. Routing protocols in 6LoWPAN

AODV has been considered as a strong candidate for 6LoWPAN due to its simplicity in finding routes [13]. Some modifications are required in AODV to suit in to 6LoWPAN environments. LOAD [6] enhances the AODV protocol according to 6LoWPAN requirements.

Several restrictions of 6LoWPAN networks include its limited energy supply, computing power and bandwidth of the wireless links connecting the 6LoWPAN nodes [9]. There are some of the routing challenges and design issues that affect the routing process in 6LoWPAN. In a multi-hop 6LoWPAN, each node can act as a data sender or receiver. Sensor nodes failure due to lack of power, physical damage, environment interference may result in rerouting or reorganisation of the network.

A 6LoWPAN routing protocol based on AODV is proposed in [6]. The proposed LOAD reduces the implementation complexity and provides load balancing in the network as compared to AODV. It maintains the routing table and route request table that are used only during route discovery phase. LOAD does not use the precursor list of AODV because Route error (RERR) message is sent only to the source. Further the protocol does not use the destination sequence number which results in reduction of size of packets and simplifies the route discovery process. The reply for the Route request (RREQ) message is sent only from the destination node which ensures loop free condition. LOAD uses the Link Quality Indicator (LQI) of 6LoWPAN MAC layer as a routing cost metric to determine the strongest route. It uses ACK message to ensure guaranteed delivery of packets.

## 2.1 MLOAD:

MLOAD stands for Multipath 6LoWPAN ad-hoc On-demand distance vector routing protocol [7]. The limitation of LOAD identified is increase in energy consumption by repeated broadcast of RREQ for route discovery process. MLOAD is proposed to reduce this network overhead. MLOAD enhances the LOAD by implementing the Ad hoc-on-demand multipath distance vector routing (AOMDV) on LOAD, to find multipath routes during route discovery process.

## 2.2 Hi-Low:

Hi-Low stands for hierarchical routing in 6LoWPAN [8].It is proposed to increase the network scalability. One of the distinctive features of 6LoWPAN is the assignment of 16-bit short addresses to the IEEE 802.15.4 devices. HiLow uses 16-bit short addresses as interface identifier for memory saving and larger scalability. HiLow exhibits parent –child architecture by initiating scanning procedures. Each node in the network discovers its parent by sending a broadcast signal. If it finds a 6LoWPAN parent node within its Personal Operating Space (POS), then it gets associated with the 6LoWPAN parent node using 16-bit short address, else it configures itself as a coordinator (parent). Every child node in the network receives its 16 bit short address from the 6LoWPAN parent node provided the following rule is satisfied.

$$\text{If } 0 < N \leq MC \tag{2.1}$$

$$\text{Then } C = MC*AP+N \tag{2.2}$$

$$\text{Else} \quad AP = 0 \tag{2.3}$$

Where

C   = address of the child node

MC = maximum number of children a parent can poses

AP = address of the Parent

N   = $n^{th}$ child node

When the current node wants to send the packet to the destination, it determines the next hop node to forward the packet. Whenever link failure is encountered, no route recovery path mechanism is performed to repair the route that carried out in LOAD. This results in unguaranteed delivery of packets in the network.

## 2.3 DYMO-Low:

DYMO-low [16] stands for dynamic MANET On-demand routing for 6LoWPAN. The DYMO-low protocol provides an effective and simple to implement routing protocol based on AODV.DYMO performs route discovery and maintenance by using RREQ, RREP and RERR messages. It operates on top of IP layer not on the link layer. DYMO protocol cannot be applied directly in 6LoWPAN routing due to its increased memory and power consumption. DYMO-low is a routing protocol proposed for 6LoWPAN which operates on link layer directly to create a mesh topology with 6LoWPAN devices, so that IP views the WPAN as the single link. DYMO-low uses both the 16-bit link layer short address and 64-bit extended address.

## 2.4 S-AODV:

S-AODV (Sink Routing Table over AODV) protocol is proposed for 6LoWPAN. This protocol provides load balancing in the network, minimise the power consumption and prolong networks lifetime. In this S-AODV protocol, the routing table is maintained only in the sink node. Sink using the routing table forwards the query packets to a specific internal node. The destined node responds to the query of the sink node through the optimal neighbouring node.

The proposed S-AODV [17] protocol consists of step-up phase and a steady state phase. Initially the sink node broadcast its status to the nodes in the network.In set-up phase every node establishes its path to the sink node through optimal neighbour node.Using this information, the sink node constructs a Sink Routing Table (SRT). In the steady-state phase, data transfer is carried out between the sink node and the destined common node. The delay and the energy consumption in the network for data forwarding is minimised by adopting this mechanism.

## 2.5 Step Parent Node (SPN):

A new path recovery algorithm called Step Parent Node (SPN) algorithm is proposed to the existing HiLow protocol. In SPN [18] algorithm every node knows its MC (Maximum Child) value (MC=4). When there is a link break the child node of the failure parent node will broadcast a step parent request message to the neighbouring nodes. The neighbouring node which has the existing number of child nodes that is less than its MC value will unicast a step parent node reply to the request sender. If the requesting node receives more reply messages then it will check the address and the Path Quality Indication (PQI) of the various senders and it will associate with the neighbouring node that has high PQI and is not the descending node of the sender.

After association the neighbour node will become the new parent node of the child node from the failure node. In this algorithm only 16-bit short addresses are used to improve the network scalability. Path recovery mechanism in conventional HiLow is solved by applying SPN algorithm in HiLow. This algorithm provides a sustainable connection along the 6LoWPAN routing path.

## 2.6 I-HiLow (Improved HiLow):

The efficiency of routing is increased by improved Hierarchical routing. In this improved Hilow the current node can acquire the information about its neighbouring nodes by broadcasting "Hello" messages in its Personal Operating Space (POS). After receiving a packet the current node "C" calculates its Parent address using the equation given below [19]:

$$AP= [(AC-1) /MC] \qquad (2.4)$$

Where

AP = address of the parent node

AC = address of the current node

MC = maximum number of children allowed

When a packet is received by a current node C, it checks for the destination 'D'. If it is the destined node 'D', the packet is delivered to the upper layer. If not, it checks to see its descendants or ascendant by using the condition discussed below.

      If C is a member of SA
      Then next hop node is AA (DC+1, D)
      If C is a member of SD
      Then next hop node is AA (DC-1, C)
      Otherwise, the next hop nodes is AA (DC-1,C)

Where

      C    = Current node

      D    = destination node

      AD = address of the destination node

      SA = set of ascendant nodes of the destination node

      SD = set of descendant nodes of the destination node

      AA (D, k) = address of the ascendant node of depth D of the node k

      DD = depth of the destination node

      DC = depth of the current node

Compared to the existing hierarchical routing, improved hierarchical routing takes minimum hop counts to reach its destination. This scheme reduces the hop-counts for communication between a node and a nearby node.

## 2.7 The bias tree avoiding technique for Hierarchical routing protocol for 6LoWPAN (TA-HiLow):

The hierarchical routing protocol is well known for the light-weight address allocation and addressing scheme. It was designed to establish a hierarchical tree with parent and child nodes to transmit packets. The problems present in the existing hierarchical routing protocol were only the address allocation and routing mechanism .But in this TA-HiLow [20], a mechanism was suggested to avoid the bias routing tree that could happen if the child nodes does not attach to the parent nodes evenly. The bias routing tree problem is avoided by transmitting attached child number information.

## 2.8 SPEED Routing Protocol in 6LoWPAN Networks:

SPEED protocol was designed for providing soft real time communication in6LoWPAN networks. In this mechanism of geographic location of packet forwarding requires that each node in the network is georeferenced. A packet is sent to the destination identified by its geographic position and global address. The destination area is identified by its centre position and radius. In this mechanism all the packets are sent towards the destination using the shortest path.SPEED supports for soft real-time,load balancing and flow shaping mechanisms making itself an effective solution in supporting packet routing in 6LoWPAN networks [21].

The limitation of LOAD include its increase in energy consumption by repeated broadcast of RREQ for route discovery process .Also when there is a link failure indicated by the RERR message, again the route has to be established since there involves only one source. ELBRP is proposed to reduce this network overhead where more than one source will be involved to establish connection with a Sink or Edge Router(ER). So even when there is a link failure the next best path would be based on the best LQI and max(distance) between nodes will be taken to route the information to the sink.

## 3. PROPOSED ELBRP (ENHANCED LOCATION BASED ROUTING OVERVIEW):

The main challenge of the Location-based routing is to integrate the concept of physical location as well as the distance information of nodes in the 6LoWPAN network. The topology of location based routing is illustrated in the Figure 4 below [11].

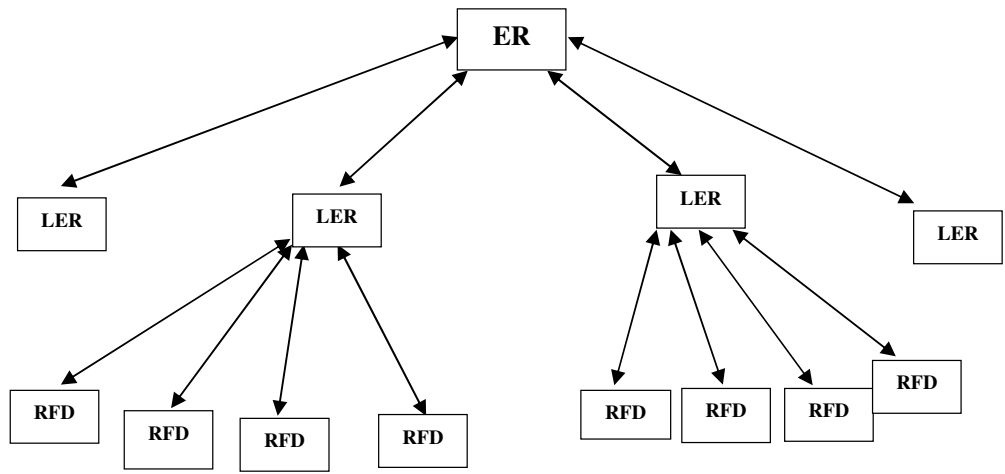

Figure 4. Topology of Location- based routing

The transmission of IPv6 packets over IEEE 802.15.4(6LoWPAN) is an architecture proposed by Internet Engineering Task Force (IETF). This IETF working group (WG) enables to carry IPv6 packets on top of the wireless personal area network (WPAN) called 6LoWPAN standardized by the IEEE 802.15.4. An adaptation layer is introduced above the MAC layer which performs mechanisms like header compression, fragmentation and reassembly to achieve this objective. A typical 6LoWPAN header stack [10] is shown below in Figure 5.

| 802.15.4 Header | Mesh Addressing Header | Fragment Header | IPv6Header Compression | IPv6Payload |
|---|---|---|---|---|
| | 4-5 bytes | 4-5 bytes | 7 bytes | 81 bytes |

Figure 5. 6LoWPAN header stack

When the payload is too large to fit in a single IEEE 802.15.4 frame, fragment header is preferred. The Mesh addressing header is used when the packet has to be transmitted over multiple radio hops and it supports layer 2 forwarding. The mesh header is ignored when 6LoWPAN frames are delivered over a single radio hop.

A stateless compression optimised for link local communication is defined in RFC 4944. The header compression mechanism reduces the IPv6 header to 3-5 bytes. 6LoWPAN routing protocol with the best link quality and maximum distance between nodes as a routing metric enables reduction in transmission overhead.

Consider a scalable 6LoWPAN network with M nodes that are deployed in an n X n terrain. The location, distance and LQI information of these M nodes are expressed as below.

$$X = x_1, x_2 \ldots\ldots\ldots\ldots x_m \tag{3.1}$$

$$Y = y_1, y_2 \ldots\ldots\ldots\ldots\ldots\ldots\ldots y_m. \tag{3.2}$$

$$\text{Distance (D)} = \text{Sqrt}(x_1 - x_2)^2 + (y_1 - y_2)^2. \tag{3.3}$$

The deployed 'M' nodes are categorised as Edge Router (ER), Local Edge Router (LER) and ordinary node or RFD (Reduced Function Device). In the proposed Enhanced Location- based routing protocol, the routing metric R (LER $_{best}$) is a product of $d_{max}$ and LQI$_{best}$ which is given as

$$R = (d_{max}) \times (LQI_{best}) \tag{3.4}$$

The architecture of the proposed ELBRP is shown in the Figure 6.

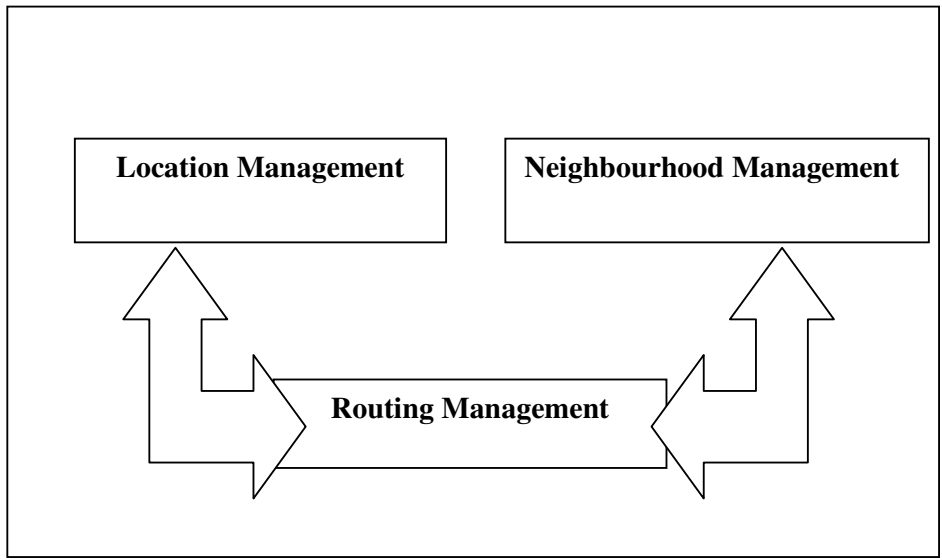

Figure 6. Architecture of the proposed ELBRP

The proposed ELBRP protocol consists of three main modules. These modules coordinate with each other to perform the task of multi-hop routing for 6LoWPAN.

The modules are 1. Location management

2. Neighbourhood management

3. Routing management

## 3.1 Location management:

During the network initialisation the ER or sink node broadcasts its location information (x, y) along with its address. The LER and other RFD nodes uses the location information of the ER for further forwarding of data along with distance and the LQI as routing metric.

## 3.2 Neighbourhood management:

Neighbourhood management consists of the neighbour discovery module. This module is triggered, when the LER send the RREP in a unicast manner. The source sends the RREQ after sensing the event .And only the nearby LERs reply for that nodes and the neighbour table information is filled in the source node.  Based on the maximum distance between LERs the best LER is chosen. Only the LER node does the RREP not the other nodes.

$D_{max}$= maximum distance (LER$_1$, LER$_2$ .......LER$_n$).

D (N, LER1) = distance between the RFD node and the LER.

Where 'N' represents the node.

The format of the RREQ packet is shown below in the Figure 7.

| TYPE | D | O | RESERVED |
|------|---|---|----------|
| LINK LAYER DESTINATION ADDRESS | | | |
| LINK LAYER DESTINATION LOCATION (x,y) | | | |
| LINK LAYER ORIGINATOR ADDRESS | | | |
| LINK LAYER ORIGINATOR LOCATION (X,Y) | | | |

Figure 7. Route Request (RREQ) PACKET FORMAT

**Table 1. Description of field Destination (D) and Originator (O)**

| Field | Value |
|-------|-------|
| D =0 | 16 bit destination address |
| D =1 | 64 bit destination address |
| O =0 | 16 bit destination address |
| O =1 | 64 bit destination address |

## 3.3    Routing Management:

Every node maintains one routing table and neighbour table. Routing table consists of the ER address, ER location, Source address and Source location. The Neighbour table consists of the address of LER, its location and LQI. The neighbour table information is filled during the RREP process of LERs. The optimal forwarding LER node is selected based on the maximum weighted value in terms of LQI and distance (d) as expressed in the equation (3.5).

$$LER_{best} = fn \{(d_{max}) \ X \ (LQI_{best})\} \qquad (3.5)$$

## 3.4 Energy Management:

In order to increase the network lifetime, the energy consumption in each node needs to be kept minimal. In the neighbour discovery process of the proposed protocol the LER nodes are alone involved in sending replies while other nodes are switched to sleep state. As a result reduction in energy consumption is achieved.

The flow diagram of the proposed ELBRP is shown in the Figure 8.

## 4.    RESULTS AND DISCUSSIONS OF THE ENHANCED LOCATION-BASED ROUTING:

The proposed ELBRP is simulated using the Network Simulator version 2.32 with IEEE 802.15.4 MAC/PHY layer support. The performance of the proposed ELBRP has been compared with LOAD in terms of PDR, Average end-to-end delay, control overhead and energy consumption in the 6LoWPAN network respectively.

**Performance Metrics:**
Control overhead: The total number of RREQ/RREP packets sent in the network for a data packet to reach the destination.

**Average End-to-End delay:**
End-to-End delay is defined as the time taken by the data packet to reach the destination node. This metric is calculated by taking the average of delays experienced by the packet received at the destination.

**Packet Delay Ratio (PDR):**
PDR is calculated as the ratio of the number of packets received at the destination node to the total number of data packets transmitted by the source node. It defines the reliability of data delivery.

**Hopcount:**
The total number of hops required to forward the data packets from source node to destination node.

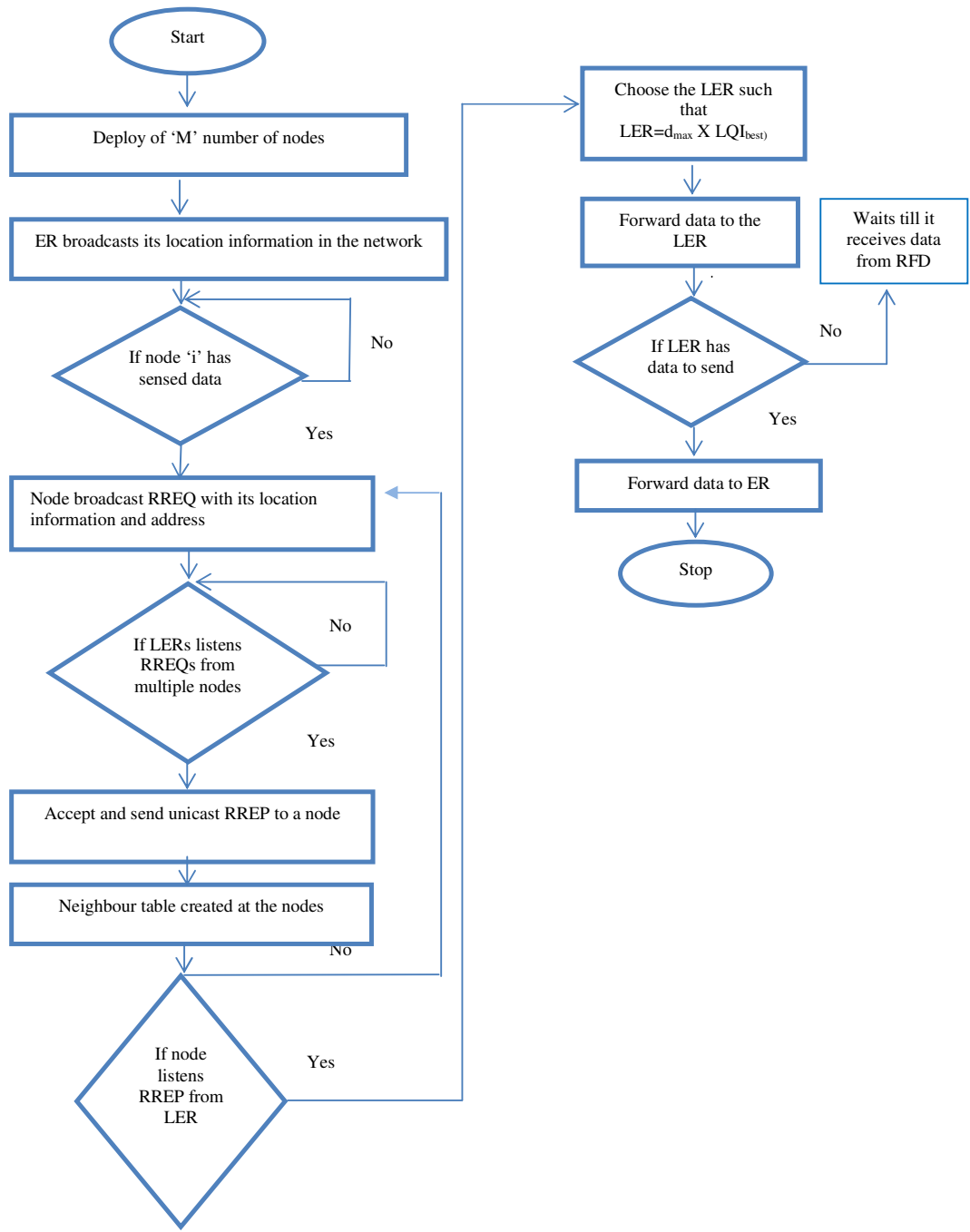

Figure 8. Flow Chart of the proposed ELBRP protocol

Consider a n X n variable size terrain with the following simulation parameters.

**Table 4.1 System parameters for simulation environment**

| Parameter | Value |
|---|---|
| Propagation Model | Two ray ground |
| MAC Type | 802.15.4 |
| Operation Mode | Non Beacon |
| Initial Energy | 1 Joules |
| Tx Power/Rx Power | 0.02mw  tx /0.01mw   rx |
| Transport Layer | UDP |
| Traffic Type | CBR |
| Packet Rate | 5 packets/sec |
| Simulation Time | 500 sec |
| Operation mode | Non Beacon |

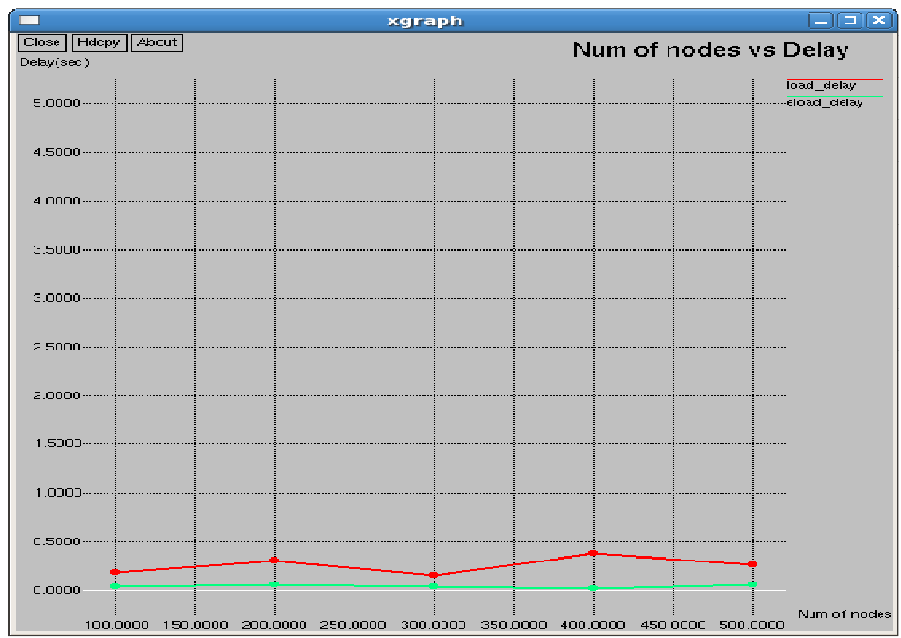

Figure 9.  Average End-to-End delay of ELBRP

Figure 9 shows the Average end-to-end delay result of the proposed ELBRP. The Average end-to-end delay of the proposed ELBRP is compared with existing LOAD. The packet takes lesser end-to-end delay as the nodes that offer maximum progresses towards destination node are selected.

Figure 10 shows the packet delivery ratio (PDR) of the proposed ELBRP.ELBRP has routing overhead, but is still nearly half that of LOAD. The high normalized routing load in LOAD is due to the broadcast of RREQ and hello messages by every node. It is found that ELBRP exhibits good PDR as this scheme selects the forwarding neighbours based on the best Link and distance from the set of LER's. The LOAD protocol exhibits lower packet delivery ratio, as this scheme selects the forwarding node which ignores the link quality issue.

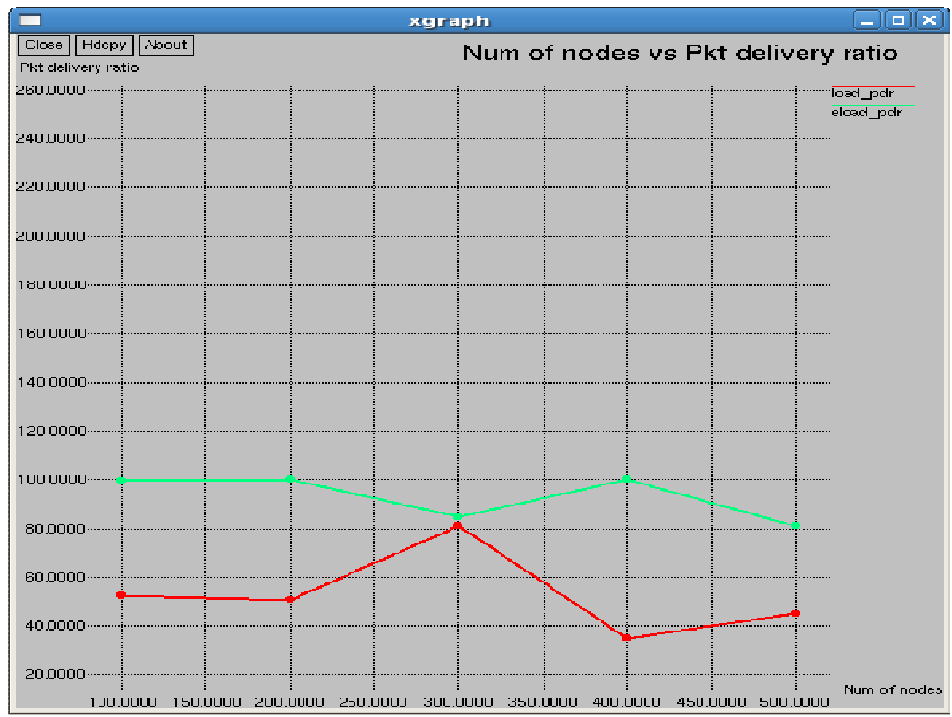

Figure 10. Packet delivery ratio

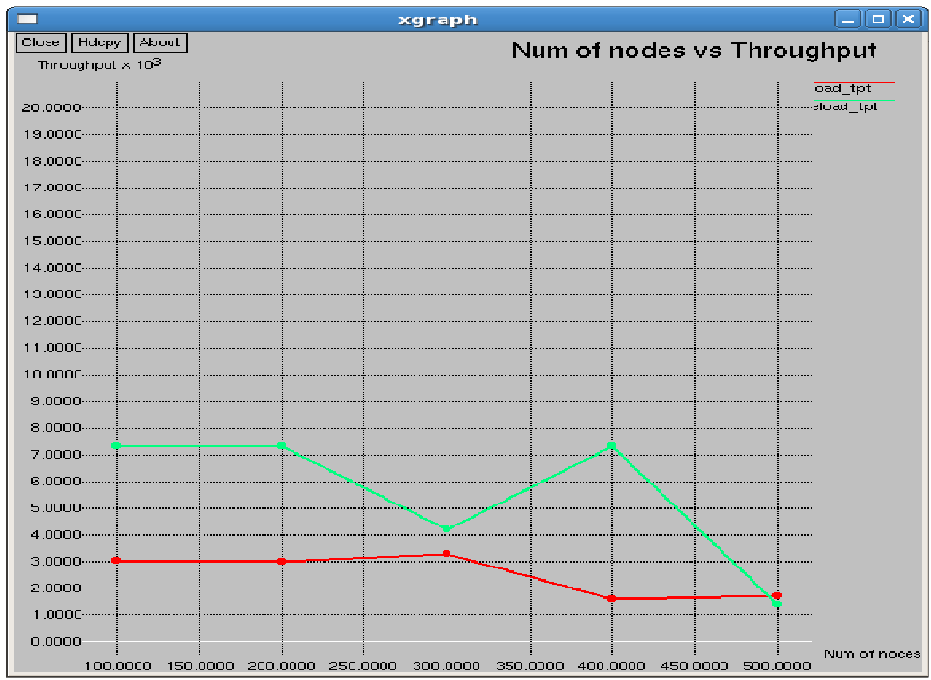

Figure 11. Throughput of ELBRP

# 5. CONCLUSION AND FUTURE WORK:

The limitations of LOAD include its repeated broadcast of RREQ for route discovery which increases the energy consumption in the node. During the link break if a local repairing fails reestablishment is done again between the source and destination to forward the data. The proposed ELBRP solves this problem by selecting a link for data forwarding based on the best link quality and the maximum distance between the nodes. Optimal path is chosen based on the maximum distance between nodes. So if the link quality and the distance are maximal the data is forwarded. This method reduces average energy consumed by the nodes in which the RREP is done only by the LER alone. The proposed protocol is simulated using the NS 2.32 simulator and the results show that the proposed protocol outperforms the existing LOAD protocol.

# REFERENCES

[1]    Jonathan Hui, David Culler, Samitha Chakrabarthi (2009) "6LoWPAN: Incorporating IEEE 802.15.4 into the IP architecture", IPSO, White paper 3.

[2]    G. Montenegro, N. Kushalnagar, J. Hui, D.Culler, (2007) "Transmission of IPv6 Packets over IEEE 802.15.4 Networks", IETF, RFC 4944.

[3]    Xin Ma, Wei Luo (2008) "The Analysis of 6LoWPAN technology", IEEE Computer Society, pp963-966

[4]    Aminulhaque, Muhammad Ikram, Hyon-Soo Cha (2009) "Route-over  vs Mesh-under Routing in 6LoWPAN",ACM.

[5]     Gee Keng Ee, CheeK yun Ng,, Nor Kamariah Noordin, Borhanudin Mohd Ali (2010) "A Review of 6LoWPAN Routing Protocols", proceeding of Asia Pacific Advanced Network.

[6]    K. Kim, Daniel park, Montenegro, N.Kushalnagar (2007) "6LoWPAN Ad hoc On-Demand Distance vector Routing (LOAD)", IETF, draft-daniel-6lowpan-load-adhoc-routing-03.txt.

[7]    Jian Ming Chang, Ting-Yun chi, Hsin-Yun Yang, Han Chieh Chao (2010) "The 6LoWPAN Adhoc On-demand Distance Vector Routing with Multi-Path scheme (MLOAD)", IET.

[8]    K.Kim, S.Yoo, J. Lee, G. Mulligan (2007) "Hierarchical Routing over 6LoWPAN (HiLow)", IETF,draft-daniel-6lowpan-hilow-hierarchical-routing-01.txt.

[9]    E. Kim, D.Kaspar, C. Gomez, Bormann "6LoWPAN Routing Requirements", IETF, draft-ietf-6lowpan-routing-requirements-04.

[10]   Jonathan W.Hui, David E.Culler, (2008) "Extending IP to Low-Power, Wireless Personal Area Networks, IEEE Internet Computing, Vol. 10, pp 37-45.

[11]   Rong DING, Haiying DU, (2011) "Location-based IP addressing in IP-enable Wireless Sensor Network", IEEE conference.

[12]   S. Thomson, T. Narten, "IPv6 Stateless Address Auto configuration", IETF draft.

[13]   C.Perkins, E.Beding Royer, S.Das (2003) "Ad hoc On-Demand Distance Vector", IETF, RFC 3561.

[14]   N.Kushalnagar, G. Montenegro, C. Schumacher (2007) "IPv6 over Low-Power Wireless Personal Area Networks (6LoWPANs): Overview, Assumptions, Problem Statement and Goals", IETF, RFC 4919.

[15]   Lamia Chaari and Lotfi Kamoun (2011) "Performance analysis of IEEE 802.15.4/ Zigbee standard under real time constraints", International Journal of Computer Networks and Communication (IJCNC), Vol.3, No.5, pp.no.235-251.

[16]    K.Kim, G.Montenegro, S.Park,   I.Chakeres, C.Perkins (2007)"Dynamic MANET On-demand for 6LoWPAN (DYMO-low)  Routing",IETF,draft-montenegro-6lowpan-dymo-low-routing-03.

[17]    Zhongyu Cao, Gang Lu (2010)"S-AODV: Sink Routing Table over AODV Routing Protocol for 6LoWPAN",Second international conference on Networks Security, pp.no 340-343.

[18]    Gee KengEe, Chee Kyun Ng, Nor Kamariah Noordin, and Borhanuddin Mohd. Ali (2010)" Path Recovery Mechanism in 6LoWPAN Routing", IEEE computer and communication engineering, pp 1-5.

[19]    Hong Yu, Jingsha He (2011) "Improved Hierarchical Routing 0ver 6LoWPAN", IEEE, pp 377-380.

[20]    Hun-Jung Lim, Tai-Myoung Chung (2009), "The Bias Routing Tree Avoiding Technique for Hierarchical Routing Protocol over 6LoWPAN", IEEE computer society, pp 232-235.

[21]    Stefano Bochhino, MatteoPetracca, Paolo Pagano, Marco Ghibaudi and Francesco Lertora (2011),"SPEED Routing Protocol in 6LoWPAN Networks ", IEEE ETFA  conference 2011.